%% file: main.tex
\documentclass[acmtog]{acmart}

\acmSubmissionID{1405}

\usepackage{booktabs} 

\usepackage{makecell}
\usepackage{multirow}
\usepackage{colortbl}
\usepackage[capitalise]{cleveref}

\copyrightyear{2025}
\acmYear{2025}
\setcopyright{cc}
\setcctype{by}
\acmConference[SA Conference Papers '25]{SIGGRAPH Asia 2025 Conference Papers}{December 15--18, 2025}{Hong Kong, Hong Kong}
\acmBooktitle{SIGGRAPH Asia 2025 Conference Papers (SA Conference Papers '25), December 15--18, 2025, Hong Kong, Hong Kong}\acmDOI{10.1145/3757377.3763871}
\acmISBN{979-8-4007-2137-3/2025/12}

\begin{document}
\title{VideoFrom3D: 3D Scene Video Generation via Complementary Image and Video Diffusion Models}

\author{Geonung Kim}
\affiliation{
  \institution{POSTECH}
  \country{Republic of Korea}
}
\email{k2woong92@postech.ac.kr}

\author{Janghyeok Han}
\affiliation{
  \institution{POSTECH}
  \country{Republic of Korea}
}
\email{hjh9902@postech.ac.kr}

\author{Sunghyun Cho}
\affiliation{
  \institution{POSTECH}
  \country{Republic of Korea}
}
\email{s.cho@postech.ac.kr}



\begin{abstract}
In this paper, we propose \methodname, a novel framework for synthesizing high-quality 3D scene videos from coarse geometry, a camera trajectory, and a reference image.
Our approach streamlines the 3D graphic design workflow, enabling flexible design exploration and rapid production of deliverables.
A straightforward approach to synthesizing a video from coarse geometry might condition a video diffusion model on geometric structure.
However, existing video diffusion models struggle to generate high-fidelity results for complex scenes due to the difficulty of jointly modeling visual quality, motion, and temporal consistency. To address this, we propose a generative framework that leverages the complementary strengths of image and video diffusion models. 
Specifically, our framework consists of a Sparse Anchor-view Generation (SAG) and a Geometry-guided Generative Inbetweening (GGI) module.
The SAG module generates high-quality, cross-view consistent anchor views using an image diffusion model, aided by Sparse Appearance-guided Sampling. 
Building on these anchor views, GGI module faithfully interpolates intermediate frames using a video diffusion model, enhanced by flow-based camera control and structural guidance. 
Notably, both modules operate without any paired dataset of 3D scene models and natural images, which is extremely difficult to obtain.
Comprehensive experiments show that our method produces high-quality, style-consistent scene videos under diverse and challenging scenarios, outperforming simple and extended baselines.
Code is available at \href{https://github.com/KIMGEONUNG/VideoFrom3D}{\textcolor{purple}{github.com/KIMGEONUNG/VideoFrom3D}}.
\end{abstract}

%
%


\input{macro.tex}

\begin{teaserfigure}
   \centering
   \includegraphics[width=\textwidth]{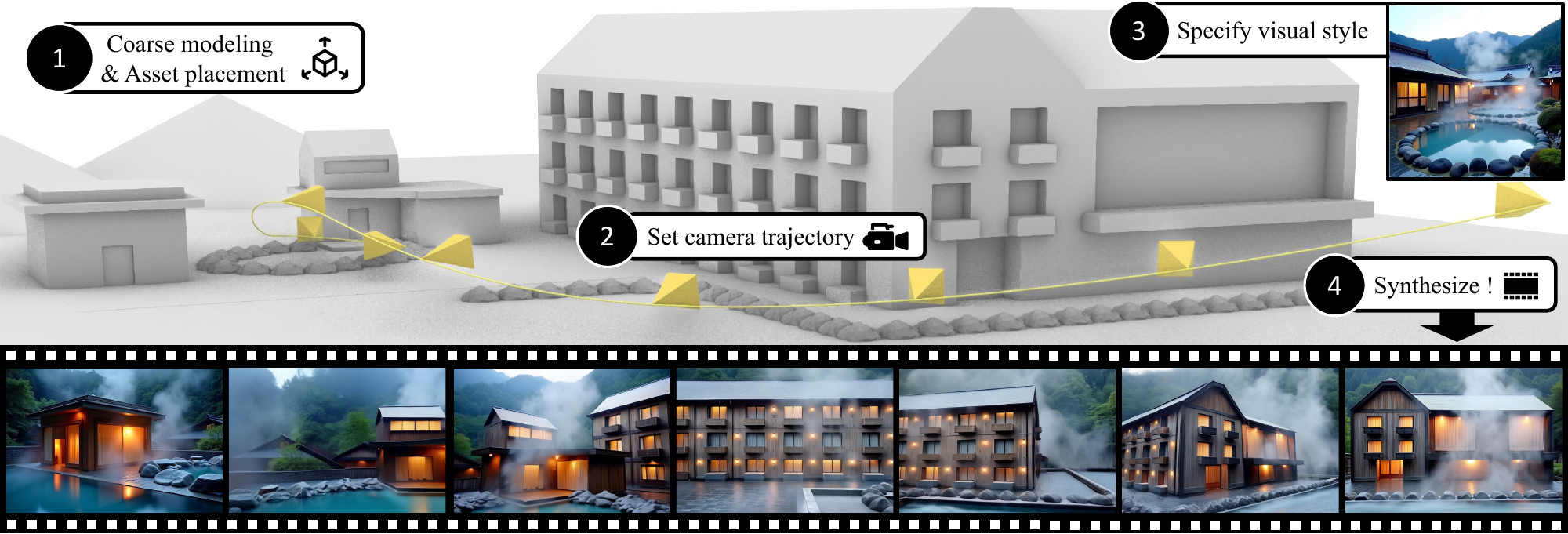}
   \caption[teaser]{
Overall framework.
(1) Users construct a scene using coarse geometry or 3D assets.
(2) A camera trajectory and (3) a reference image are provided.
(4) The framework then generates a high-quality video reflecting the specified style, structure, and camera motion. 
The synthesized video sequence shows consistent, high-quality visuals that reflect the input geometry and reference style, including challenging visual elements such as rising steam.
   }
   \label{fig:teaser}
\end{teaserfigure}

\maketitle

\input{sections/a_intro.tex}

\input{sections/b_relatedwork}

\input{sections/c_method}

\input{sections/d_experiments}

\input{sections/e_conclusion}

\begin{acks}
This work was supported by Institute of Information \& communications Technology Planning \& Evaluation (IITP) grants funded by the Korea government (MSIT) (No.2019-0-01906, Artificial Intelligence Graduate School Program(POSTECH), No. RS-2024-00457882, AI Research Hub Project, No.2021-0-02068, Artificial Intelligence Innovation Hub).
\end{acks}

\bibliographystyle{ACM-Reference-Format}
\bibliography{bibliography}


\clearpage

\end{document}

%% file: macro.tex
\def\methodname{VideoFrom3D} 

\newcommand{\Eq}[1]  {Eq.\ (#1)}
\newcommand{\Eqs}[1] {Eqs.\ (#1)}
\newcommand{\Fig}[1] {Fig.\ #1}
\newcommand{\Figs}[1]{Figs.\ #1}
\newcommand{\Tbl}[1]  {Tab.\ #1}
\newcommand{\Tbls}[1] {Tabs.\ #1}
\newcommand{\Sec}[1] {Sec.\ #1}
\newcommand{\SSec}[1] {Sec.\ #1}
\newcommand{\Secs}[1] {Secs.\ #1}
\newcommand{\Alg}[1] {Alg.\ #1}

\newcommand{\setone}[1] {\left\{ #1 \right\}} 
\newcommand{\settwo}[2] {\left\{ #1 \mid #2 \right\}} 

\newcommand{\kkw}[1]{{\textcolor[rgb]{0.0,0.6,0.4}{#1}}}
\newcommand{\sunghyun}[1]{{\textcolor[rgb]{0.6,0.0,0.6}{sunghyun: #1}}}

\newcommand{\bb}[1]{\textbf{\textit{#1}}}
\newcommand{\xx}{\textcolor{red}{XX}}
\newcommand{\fix}[1]{\textcolor{red}{#1}}
\newcommand{\rt}[1]{\textcolor{red}{#1}}
\newcommand{\edit}[1]{\textcolor{red}{#1}}
\newcommand{\bt}[1]{\textcolor{blue}{#1}}
\newcommand{\idea}[1]{\textcolor{olive}{IDEA: #1}}

\newcommand\Myperm[2][^n]{\prescript{#1\mkern-2.5mu}{}P_{#2}}
\newcommand\Mycomb[2][^n]{\prescript{#1\mkern-0.5mu}{}C_{#2}}






%% file: sections/a_intro.tex
\section{Introduction} 
\label{sec:Introduction}
3D graphic design refers to the process of creating visually compelling three-dimensional representations for communication, simulation, or artistic purposes. It serves diverse purposes across domains, including conveying design intent in architecture, building immersive worlds in games, generating photorealistic effects in film, and enabling real-time interaction in VR and metaverse applications. Across these domains, the underlying production process typically follows a common sequence of stages. The design workflow usually begins with a concept development phase, where rough visual ideas, a preliminary 3D scene layout, and a camera trajectory are established. This is followed by detailed production steps including modeling, texturing, and lighting, with a focus on the regions that will be visible in the final render. The process culminates in the rendering stage, which produces the final visual output such as images or videos \cite{bettis2005digital,hamdani20233d}.

In practice, however, this workflow does not proceed in a single pass but involves repeated iterations across stages. Specifically, to refine the design, designers often receive feedback after the rendering stage from clients or collaborators, and return to earlier stages to revise the work. One major challenge is that even minor changes in a single component can require extensive adjustments in multiple stages of the workflow~\cite{bettis2005digital,lord2024openusd}. For example, when the intended camera trajectory or scene composition is modified, previously detailed modeling, texturing, and lighting may all require updates, as the regions visible to the camera also change. Similarly, changes in the visual concept often necessitate broad adjustments in both texturing and lighting. Because each stage is time-consuming and requires a high level of expertise, even minor revisions can result in significant increases in production cost.  


In this paper, we propose \methodname, a novel framework for synthesizing high-quality 3D scene videos from coarse geometry.
By leveraging generative models, our approach streamlines the 3D graphic design pipeline, offering a faster and more flexible alternative to the traditional, labor-intensive workflow.
\cref{fig:teaser} illustrates our framework step by step. Firstly, (1) a user constructs a scene by modeling coarse geometry or by assembling a scene using pre-existing 3D assets. 
Then, (2) a camera trajectory and (3) a reference image representing the desired visual concept are provided.
Given these inputs, (4) our framework synthesizes a high-quality video that reflects the specified style, structure, and camera motion through a generative process. 
This addresses the aforementioned inefficiencies in two ways. First, by relying on coarse modeling and asset placement instead of detailed modeling, the framework allows flexible adaptation to changes in scene layout and concept. Second, the generative synthesis strategy based on a reference style enables efficient adaptation to changes in visual style or camera trajectory without redoing time-consuming texturing and lighting. Consequently, our framework can be employed in early-stage design development by enabling rapid iteration and visual exploration prior to labor-intensive asset production. Alternatively, for visualization-only purposes, the generated output can serve directly as the final deliverable. 

\input{figures/image-video-cmp}
Based on the recent success of video diffusion models in 3D scene generation, a na\"ive solution to this problem would be to condition a video diffusion model on geometric information, such as with a depth-based ControlNet~\cite{controlnet}. However, this na\"ive solution faces a core limitation: \textit{video diffusion models are fundamentally limited in handling complex scenes compared to image diffusion models}. \cref{fig:image-video-cmp} compares the outputs of image and video diffusion models on a complex scene. While the image model produces realistic building details, the video models generate distorted structures with lower visual quality, despite having far more parameters.
This limitation is also evident in the quantitative comparison in \cref{tab:idm_vdm}.
The primary reason lies in the inherent challenges of video synthesis.
Unlike image diffusion models, which focus exclusively on generating high-quality still frames, video models must simultaneously learn to synthesize individual frames, ensure realistic motion, and maintain temporal coherence across video frames.
This added complexity makes it harder for video diffusion models to match the individual image quality achieved by image diffusion models.


\input{tables/cmp_idm_vdm}

To address this issue, we leverage the complementary strengths of image and video diffusion models.
Specifically, image diffusion models are highly effective at generating high-quality frames with fine spatial detail, while video diffusion models excel at maintaining temporal consistency across sequences.
Building on this insight, our key idea is to first generate a set of high-quality, multi-view-consistent anchor frames using an image diffusion model, and then interpolate the anchor frames using a video diffusion model to synthesize temporally coherent intermediate frames, instead of synthesizing complex scenes from scratch.
To realize this, we introduce two key modules: a Sparse Anchor-view Generation (SAG) module and a Geometry-guided Generative Inbetweening (GGI) module.
The SAG module produces high-quality, multi-view-consistent anchor views using an image diffusion model. A major challenge at this stage is preserving multi-view consistency.
To overcome this, we introduce Sparse Appearance-guided Sampling, which adopts a distribution alignment strategy and leverages appearance guidance from a warped adjacent view to generate consistent results.
Interpolating these anchor views, the GGI module generates consistent intermediate frames using a video diffusion model. To ensure natural interpolation and precise trajectory alignment, we incorporate flow-based camera control and structural guidance into the GGI module.
Notably, our modules achieve high visual quality without relying on any paired dataset of 3D scene models and natural images, which are typically unavailable in practice.

\input{figures/pipeline}

A potential alternative to generate a video from the input geometry is to synthesize textures for a mesh model using recent texturing techniques~\cite{texpainter,texgen,texture,text2tex,paint3d}, and then render the result.
However, our method differs in two important ways.
First, texturing-based approaches require detailed geometry to produce natural-looking results. When applied to coarse geometry, this often leads to visual artifacts, e.g., flowers or grass appearing unnaturally flattened onto planar ground surfaces.
Second, texture maps are inherently static and cannot capture dynamic, view-dependent effects such as reflections, flickering flames, or flowing streams. By directly generating video, our method naturally models such variations. These advantages come with certain limitations: our framework does not support real-time navigation or enforce pixel-level consistency across views. Nonetheless, it offers a compelling alternative for fast, stylized scene video generation from minimal input.

Our main contributions can be summarized as follows:
\begin{itemize}
\item We propose \methodname, a novel framework that synthesizes a high-quality 3D scene video given coarse geometry, a camera trajectory, and a reference image.
\item We propose a two-stage approach that leverages the complementary strengths of image and video diffusion models, where the SAG module uses image diffusion for anchor view generation, and the GGI module applies video diffusion to interpolate the anchor views.
\item Extensive experiments show that our method robustly synthesizes high-fidelity videos under diverse and challenging scenarios, outperforming na\"ive and extended baselines.
\end{itemize}


%% file: figures/image-video-cmp.tex
\begin{figure}
    \centering
    \includegraphics[width=0.90\linewidth]{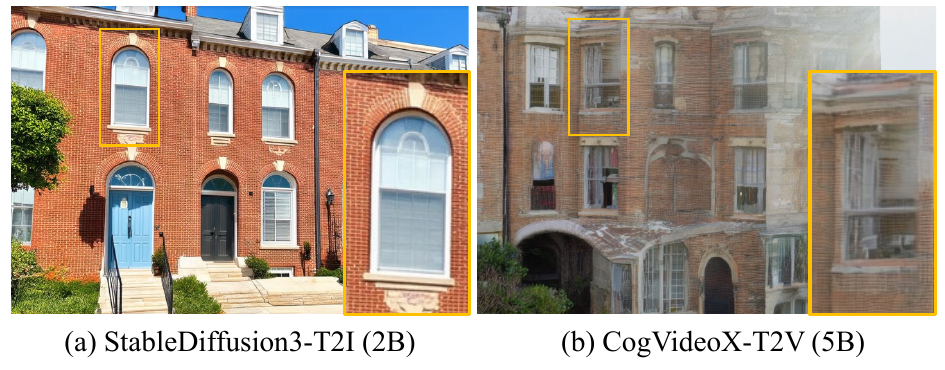}
    \vspace{-1mm}
    \caption{
    Comparison of outputs from image and video diffusion models, conditioned only on a text prompt. Model sizes are noted in parentheses.
    }
    \vspace{-6mm}
    \label{fig:image-video-cmp}
\end{figure}

%% file: tables/cmp_idm_vdm.tex
\begin{table}[t]
    \centering
    \caption{
    Image aesthetics (CLIP-A) and quality (MUSIQ) are compared across 1,000 generated samples (parameter size in parentheses). Prompts are auto-generated by GPT to describe complex outdoor scenes. For video models, only the first frame of each video is evaluated.
    }
    \scalebox{0.78}{
    \begin{tabular}{lccc}
        \toprule
        & StableDiffusion3 (2B) & CogvideoX (2B) & CogvideoX (5B)   \\
        \hline\hline
        CLIP-A$\uparrow$ & {\bf 5.942} & 5.119 & 5.144 \\
        MUSIQ$\uparrow$ & {\bf 67.04} & 56.36 & 58.20 \\
        \bottomrule
        \end{tabular}
        }
    \vspace{-2mm}
        
    \label{tab:idm_vdm}
\end{table}

        

%% file: figures/pipeline.tex
\begin{figure*}[t]
    \centering
    \includegraphics[width=1.0\linewidth]{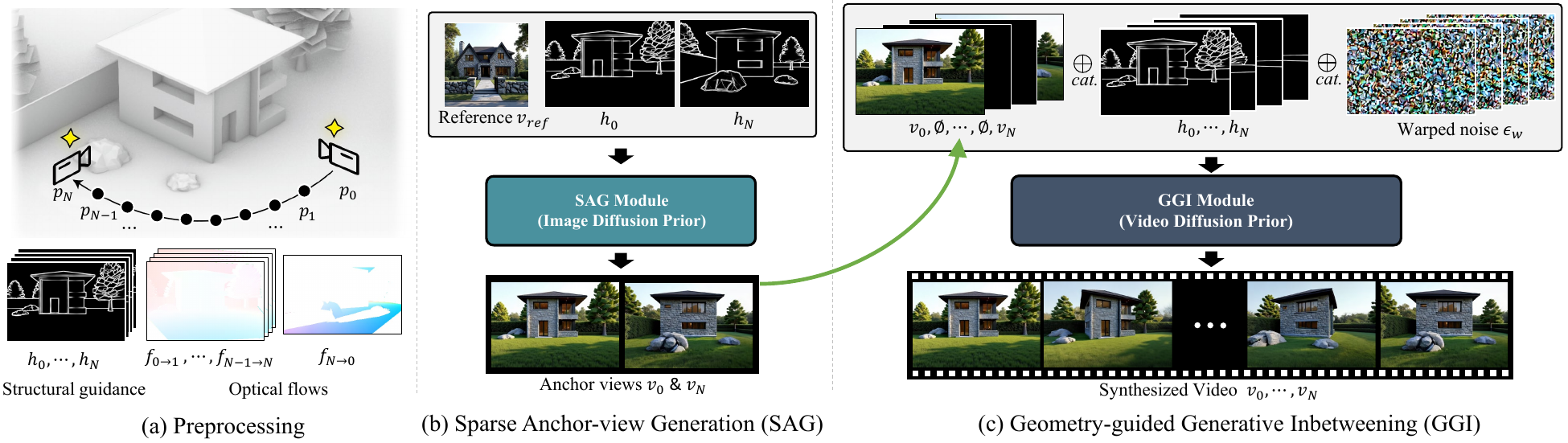}
    \caption{Overall pipeline.
(a) Preprocessing extracts structural edges and optical flows from geometry and camera trajectory.
(b) SAG module generates high-quality anchor views $v_0$ and $v_N$ using an image diffusion prior.
(c) GGI module interpolates intermediate frames with a video diffusion prior. 
    }
    \label{fig:pipeline}
\end{figure*}

%% file: sections/b_relatedwork.tex
\section{Related Work} 
\label{sec:Related Work}

\paragraph{Geometry-guided Video Generation}
Structure-conditioned video generation methods~\cite{vace,cosmos-transfer} provide a simple baseline for our problem by conditioning video outputs on depth or edge maps rendered from the input geometry.
However, their results often suffer from poor visual quality in complex scenes, due to the limited prior of video diffusion models.
Another line of work explores 3D generation methods conditioned on geometry~\cite{fantasia3d,latentnerf,coin3d,mvedit,zero123++,phidias}, whose outputs can be rendered into videos. Yet, they are limited to object-level generation.
In contrast, Urban Architect~\cite{urbanarchitect} and ControlRoom3D~\cite{controlroom3d} propose scene-scale 3D generation methods conditioned on semantic proxy geometry.
However, the former produces blurry results due to the limited guidance from the SDS-based image prior, while the latter is limited to rectangular room layouts. 
To the best of our knowledge, we are the first to enable geometry-guided generation of high-quality, large-scale, and versatile 3D scenes.

\paragraph{Few-shot 3D Reconstruction}
Few-shot 3D reconstruction recovers a 3D scene from a small number of input views.
Similar to our GGI module, recent approaches interpolate keyframes across sparse input views using video diffusion models.
To this end, most approaches first construct an intermediate point-based representation from sparse views, which is then used to condition the video diffusion model for intermediate frame synthesis~\cite{viewcrafter,mvsplat360,reconx,uni3c,see3d,gen3c}.
For example, MVSplat360~\cite{mvsplat360} builds a coarse 3D Gaussian Splatting~\cite{3dgs} via feedforward prediction to guide video generation.
However, such methods often fail to construct reliable structure under wide keyframe baselines, resulting in severe artifacts.
In other lines of work, LVSM~\cite{lvsm} performs regression-based interpolation but lacks strong generative priors, leading to failure under complex transitions.
SEVA~\cite{seva} leverages multi-view diffusion models with Plucker embeddings to condition the camera trajectory. However, this explicit pose representation suffers from scale ambiguity, making it difficult to follow the intended path, and also causes temporal flickering inherent to multi-view diffusion. 
In contrast, our GGI module leverages a video diffusion prior and structure-aware conditioning, enabling accurate, smooth, and robust interpolation.

%% file: sections/c_method.tex
\input{figures/sa-sdedit}

\section{Method} 
\label{sec:Method}



We define the \methodname~task as follows.
Let $\mathbb{M}$ be a 3D mesh model, and $v_{ref}$ be a reference image representing the desired style.
Additionally, let $\mathbb{P} = \{p_0, \dots, p_N\}$ be a camera trajectory, where $p_i$ is the camera pose at the $i$-th frame, such that $0 \le i \le N$.
Given these inputs, the goal of \methodname~is to generate a sequence of images $\mathbb{V} = \{v_0, \dots, v_N\}$ that are synthesized from the corresponding camera poses in $\mathbb{P}$, faithfully reflect the geometry $\mathbb{M}$, and are consistently stylized according to $v_{ref}$.


\cref{fig:pipeline} illustrates the overall pipeline of \methodname, which consists of three stages.
The preprocessing stage extracts structural guidance $\{h_0, \dots, h_N\}$ and optical flows $\{ f_{0\rightarrow 1}, \dots, \allowbreak f_{N-1\rightarrow N}, \allowbreak f_{N \rightarrow 0}\}$ from an input 3D model $\mathbb{M}$ and a camera trajectory $\mathbb{P}$, where $f_{i \rightarrow j}$ denotes the optical flow from view $i$ to view $j$.
The structural guidance constrains the SAG and GGI modules to synthesize images accurately reflecting the geometry of $\mathbb{M}$.
$f_{N \rightarrow 0}$ is used by the SAG module to enforce cross-view consistency between $v_0$ and $v_N$, while $\{ f_{0\rightarrow 1}, \dots, f_{N-1\rightarrow N}\}$ is used by the GGI module to provide guidance on the camera trajectory.
In the next stage, the SAG module generates high-quality anchor views, $v_0$ and $v_N$, using an image diffusion prior. The generation of $v_N$ is conditioned on $v_0$ for cross-view consistency.
Finally, the GGI module interpolates between the anchor views using a video diffusion prior, producing the full sequence $\mathbb{V}$.
To support longer camera trajectories, the pipeline can be iteratively applied by treating $v_N$ as the next starting frame $v_0$.

In the following, we describe each stage of our method and the training strategy for the GGI module in detail.

\subsection{Preprocessing}
\label{ssec:preprocessing}
To compute the optical flow $f_{i \rightarrow j}$, we first backproject a coordinate-encoded color map from $p_j$ onto the 3D mesh, where each pixel encodes its own image-space coordinate.
We then reproject the coordinate-encoded mesh onto the image plane of $p_i$, allowing us to establish dense correspondences based on color information, which are then used to derive the optical flow.

As structural guidance $h_i$, we employ a 2D edge map, which represents the shapes of the input 3D model $\mathbb{M}$ as projected at camera pose $p_i$.
Specifically, we empirically select four types of geometry-based edges: silhouette, crease, object boundary, and intersection, provided by Blender (see $h_0$ and $h_N$ in Fig. 3).
For more details on edge extraction, refer to the supplemental document.

While depth and normal maps can serve as structural guidance, they are generally less reliable compared to edge maps in preserving geometric structures.
Depth maps suffer from scale inconsistencies across different 3D models.
Even within a single scene, some geometric details, such as windows on a building’s surface, may have only slight depth differences from their surroundings.
These issues make it difficult for diffusion models to faithfully reflect the geometric guidance in depth maps.
Normal maps, while invariant to scale, are less effective at representing geometric boundaries when distinct surfaces that should be separated exhibit similar normal values.
In contrast, edge maps remain scale-invariant and precisely define object boundaries, ensuring robust shape preservation.

\subsection{Sparse Anchor-view Generation (SAG)}
\label{ssec:sag}

The SAG module synthesizes high-quality anchor views, $v_0$ and $v_N$, using FLUX-dev~\cite{flux}, a state-of-the-art text-to-image diffusion model.
For high-quality anchor view synthesis, the SAG module needs to satisfy three criteria: reflect the structural conditions $h_0$ and $h_N$, match the visual style of the reference image $v_{ref}$, and maintain cross-view consistency between $v_0$ and $v_N$. We describe how each of these criteria is addressed in the following.

To incorporate structural guidance, the SAG module adopts ControlNet~\cite{controlnet} as the conditioning mechanism.
To this end, rather than training a ControlNet specifically on our structural guidance, we adopt a pretrained ControlNet using edges from the HED edge detector~\cite{hed}, which extracts perceptually-aligned edges from 2D images\footnote{\url{https://huggingface.co/XLabs-AI/flux-controlnet-hed-v3}}.
Although HED edges do not perfectly match those in our structural guidance, we empirically found that this approach performs effectively. 
More importantly, using HED edges eliminates the need for a specialized dataset of 3D-model-derived edges paired with natural images for training, which is extremely difficult to obtain.


To incorporate the style reference, we adopt a distribution alignment strategy. Specifically, we add LoRA~\cite{lora} layers to both the image diffusion model and ControlNet, and train them using the reference image $v_{ref}$ with a unique identifier prompt before synthesizing anchor views. This strategy aligns the target distribution of the diffusion model to the reference style, enabling style-consistent anchor view generation. As a result, the start view $v_0$ is synthesized using the style-aligned diffusion model, guided by the identifier prompt and the structural condition $h_0$.

\paragraph{Sparse Appearance-guided Sampling}
To generate the end view $v_N$ while maintaining cross-view consistency with $v_0$, we propose a Sparse Appearance-guided Sampling strategy (\cref{fig:sa-sdedit}).
Our strategy first obtains a sparse observation $v_{0 \rightarrow N}$ by warping $v_0$ to the end view $v_N$ using the optical flow $f_{N \rightarrow 0}$.
The observed regions in $v_{0 \rightarrow N}$ often exhibit distortions due to excessive warping (\cref{fig:sa-sdedit}a).
Nevertheless, they still retain useful semantic and appearance information that supports cross-view consistency.
To exploit this information, we replace the latent of $v_N$ with that of $v_{0 \rightarrow N}$, in the regions observed in $v_{0 \rightarrow N}$, during the diffusion sampling process~\cite{elevate3d}.

Specifically, we first compute the latent of $v_{0\rightarrow N}$, denoted as $\bar{z}_0$, using the encoder of the image diffusion model.
Additionally, we generate a binary mask $m$ to indicate the observed regions in $v_{0 \rightarrow N}$, and obtain a downsampled version, $\bar{m}$, according to the size of the latent $\bar{z}_0$.
We then randomly initialize the latent of $v_N$, denoted as $z_{T}$, where $T$ represents the total number of diffusion timesteps.
The standard diffusion process iteratively denoises the latent $z_{t}$ from $t=T$ to $t=0$.
To guide the diffusion process to synthesize an image consistent with $v_{0 \rightarrow N}$, we perform a replacement operation before denoising at each timestep $t$, defined as:
\begin{equation}
    z_t \leftarrow \bar{m} \odot \bar{z}_t + (1 - \bar{m}) \odot z_t,
    \label{eq:replacement}
\end{equation}
where $\odot$ is element-wise multiplication.
In \cref{eq:replacement}, $\bar{z}_t$ is the latent of $v_{0 \rightarrow N}$ at timestep $t$, obtained by adding noise to $\bar{z}_0$ following the noise scheduling of the image diffusion model.
We apply the replacement operation only for early timesteps to reflect only the semantic and color information without distorted details in $v_{0\rightarrow N}$.

Through this process, we can synthesize natural-looking content across both observed and unobserved regions.
Thanks to the replacement operation, the details generated for observed areas adhere closely to the semantic structures provided by the warped image $v_{0\rightarrow N}$.
Meanwhile, the synthesis of unobserved regions maintains consistency with the visual characteristics of the observed areas, ensuring spatial coherence throughout the final output (\cref{fig:sa-sdedit}b).
We use 25 diffusion steps to generate each anchor view, and apply the replacement operation for the first 12 steps.

It is noteworthy that the proposed approach is made possible thanks to the distribution alignment using the style reference image performed before synthesizing anchor views.
Without the distribution alignment, the aforementioned approach fails to produce coherent results in observed regions.
This is because unknown regions typically occupy a much larger area than observed ones, making it difficult for the model to generate consistent content based on limited guidance.
As a result, as shown in \cref{fig:sag_A}(c), the model often produces entirely different content with noticeable seams and inconsistency across the boundary.
In contrast, the distribution alignment process narrows the solution space toward the reference style, enabling coherent synthesis, as shown in \cref{fig:sag_A}(b).
The supplemental document provides additional discussions on the difference against inpainting approaches and multi-view-diffusion-based approaches as well as a detailed pseudocode.

\input{figures/sag_A}

\paragraph{Style Variation}
Depending on the application, multiple reference styles may be required in a single scene. For example, when a scene includes transitions between indoor and outdoor areas, each region exhibits different structural and appearance characteristics, necessitating separate style references. However, training individual LoRA models for each style is cumbersome. To address this, we train a single LoRA model, assigning a unique identifier prompt to each reference image instead of training separate models for different styles. During anchor view generation, we selectively apply the desired style by using the identifier prompt corresponding to the target reference.
In another scenario, users may want to apply global style variations such as seasonal changes or tonal shifts.
In a similar vein, to avoid training additional LoRA models, we adopt a post-prompting strategy in which style variation is introduced at inference time by modifying the text prompt. Specifically, anchor views are generated using a unique identifier prompt combined with an additional style description, such as `winter' or `cozy'.

\input{figures/qual_A}

\input{figures/qual_lora_2col}
\input{figures/4dtransition}
\input{figures/hedofdepth}

\subsection{Geometry-guided Generative Inbetweening (GGI)}
\label{ssec:ggi}

The GGI module synthesizes a high-quality video frames $\mathbb{V}$ from the anchor views $v_0$ and $v_N$, by leveraging a video diffusion prior. 
To effectively perform the inbetweening task, we build upon a pretrained Image-to-Video (I2V) diffusion model, CogVideoX-5B-1.0~\cite{cogvideox}. To condition on both endpoints, we encode the start and end frames $v_0$ and $v_N$ using the VAE encoder $\mathcal{E}$. Zero-valued latents $\emptyset$ are used for the intermediate frames, resulting in ${V} = [\mathcal{E}(v_0), \emptyset, \cdots, \emptyset, \mathcal{E}(v_N)]$, where $[\cdot]$ denotes stacking along the temporal dimension. The feature $V$ is concatenated with the noisy latent along the channel dimension.
Additional implementation details on encoding the conditions with the 3D causal VAE of CogVideo-X are provided in the supplemental document.

To condition on the camera trajectory $\mathbb{P}$, we adopt a flow-based camera control approach similar to Go-with-the-Flow~\cite{gowiththeflow,flovd}. 
Specifically, we obtain a warped noise volume, denoted as $\epsilon_w$, that implicitly encodes the camera motion. To this end, we sample the initial noise for the first frame and recursively warp it using the consecutive optical flows $\{ f_{0\rightarrow 1}, \cdots, f_{N-1\rightarrow N} \}$ while preserving Gaussianity.
To reflect the motion information in the generation process, we employ the pretrained flow-aware LoRA module from Go-with-the-Flow~\shortcite{gowiththeflow}.

While the warped noise provides approximate guidance for the overall camera motion, it is insufficient for accurately capturing the intended motion trajectory, for a couple of reasons. First, the warped noise volume is constructed in a downsampled latent space, e.g., $8\times$ smaller spatially and $4\times$ temporally, inherently limiting the granularity of motion guidance. In addition, to preserve gaussianity during the noise warping process, Gaussian noise is continually re-injected, which results in the flow information being only implicitly encoded. This makes precise camera control challenging and often leads to structural distortions. To address this, we additionally concatenate the VAE-encoded HED edge maps $\mathcal{E}([h_0,\cdots,h_N])$ to the latent feature as structural guidance.

Finally, the diffusion sampling step of the GGI module is represented as:
\begin{equation}
   \epsilon_{\Theta,\pi}\left(Z_t \oplus {V} \oplus \mathcal{E}\left(\left[h_0,\cdots,h_N\right]\right),\ t\right) \mapsto Z_{t-1},
  \label{eq:inference}
\end{equation}
where $\epsilon_{\Theta,\pi}$ denotes the diffusion sampling operation with parameters $\Theta$ for the base video diffusion model and $\pi$ for the flow-aware LoRA. 
$Z_t$ is the noisy latent at timestep $t$, initialized with $\epsilon_w$, and $\oplus$ indicates channel-wise concatenation.

\subsection{GGI Module Training}
\label{ssec:training}

Training the GGI module ideally requires coarse geometry, camera trajectories, and their corresponding high-quality multi-view images, but such datasets are rarely available.
To approximate this setting, we use the DL3DV-10K~\cite{dl3dv} dataset, which provides various videos of static scenes.
Specifically, for each training video $X$, we compute the optical flows using RAFT~\cite{raft} to generate the warped noise $\epsilon_w$. For each frame, we extract the HED edge map $h_i$ for structural guidance. 

While training the GGI module requires edge maps derived from 3D models for structural guidance, the DL3DV-10K dataset lacks such 3D models.
Thus, instead of using 3D models, we synthesize edge maps from training videos as illustrated in \cref{fig:hedofdepth}.
The 3D-model-derived edge maps in our scenario exhibit two key characteristics: they contain no appearance information, such as texture, and they are derived from coarse geometry.
To replicate these properties during training, we first estimate depth maps from training videos using an off-the-shelf depth estimator~\cite{midas}, and apply the HED edge detector~\cite{hed} to the estimated depth maps.
Since depth maps inherently lack textures, and the HED detector selectively extracts strong structural contours, ignoring weak edge signals, this approach produces edge maps that closely resemble inference-time structural guidance, effectively reducing the domain gap between training and inference.
Finally, the training objective of the GGI module is defined as:
\begin{align}
\underset{\Theta}{\operatorname{argmin}} \
\mathbb{E}_{X,t}\left[
  \left\| 
  \epsilon_{w} - \epsilon_{\Theta,\pi}\left(Z_t \oplus V \oplus H,\ t\right) 
  \right\|^2
\right], \\
H = \mathcal{E}\left([\mathcal{A}_e(\mathcal{A}_d(x_0)),\ \cdots,\ \mathcal{A}_e(\mathcal{A}_d(x_N))]\right), \nonumber
\label{eq:training}
\end{align}
where $\mathcal{A}_e$ denotes the HED edge estimator, $\mathcal{A}_d$ denotes the depth estimator, and $x_i$ denotes the $i^{\text{th}}$ frame of the video $X$.

%% file: figures/sa-sdedit.tex
\begin{figure*}[t]
    \centering
    \includegraphics[width=\linewidth]{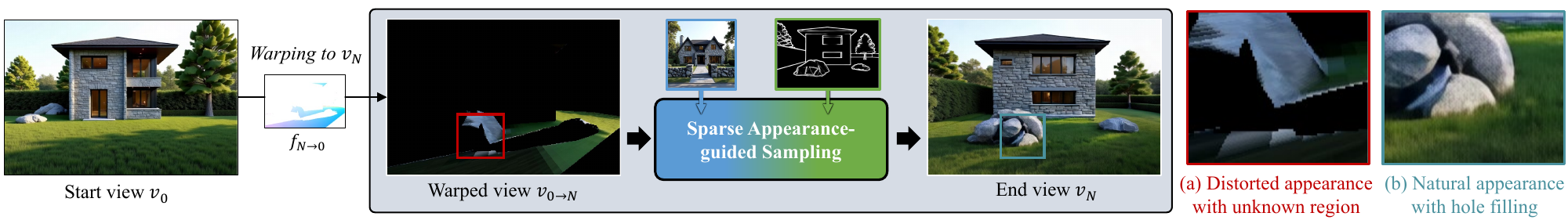}
    \caption{
To generate a multi-view coherent $v_N$ from $v_0$, Sparse Appearance-guided sampling uses the distorted appearance of the warped image as guidance during sampling, achieving the successful generation of a coherent and high-quality $v_N$.
}

    \label{fig:sa-sdedit}
\end{figure*}

%% file: figures/sag_A.tex
\begin{figure}[t]
    \centering
    \includegraphics[width=\linewidth]{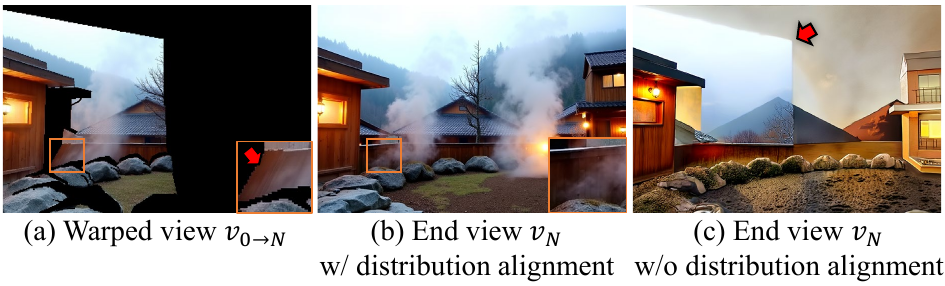}
    \vspace{-5mm}
    \caption{
    Effect of distribution alignment in generating the end view $v_N$ using Sparse Appearance-guided sampling.
    }
    \vspace{-1mm}
    \label{fig:sag_A}
\end{figure}

%% file: figures/qual_A.tex
\begin{figure*}[!t]
    \centering
    \includegraphics[width=0.87\linewidth]{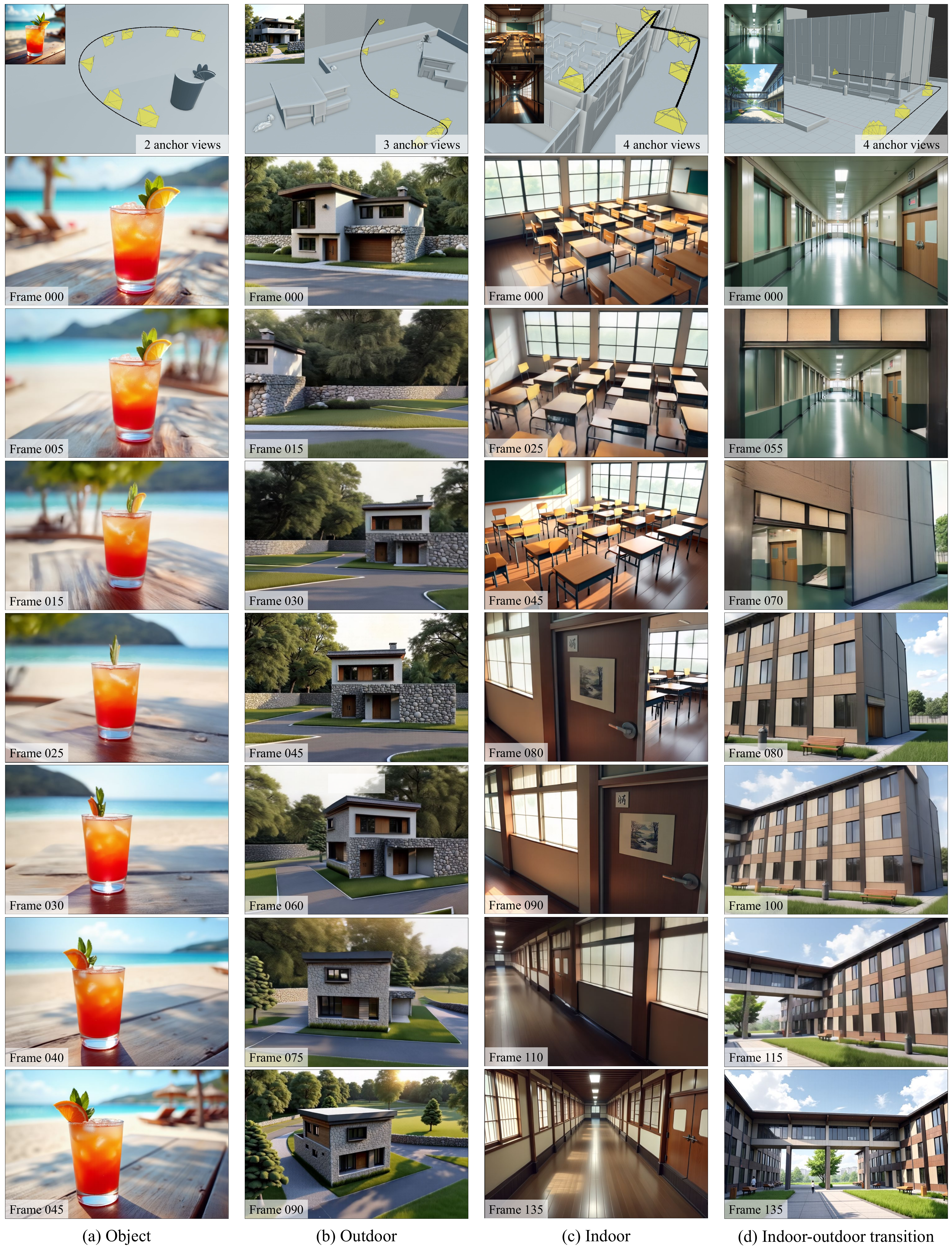}
    \vspace{-4mm}
    \caption{
    Qualitative results across various scenarios. The first row illustrates scene information: the style reference (top-left), the camera trajectory (black line), camera positions corresponding to each generated view shown below (yellow), and the number of anchor views (bottom-right). The input geometry in (c) and (d) is from TurboSquid (\textcopyright Okhey).
    }
    \label{fig:qual}
\end{figure*}

%% file: figures/qual_lora_2col.tex
\begin{figure*}[t]
    \centering
    \includegraphics[width=\linewidth]{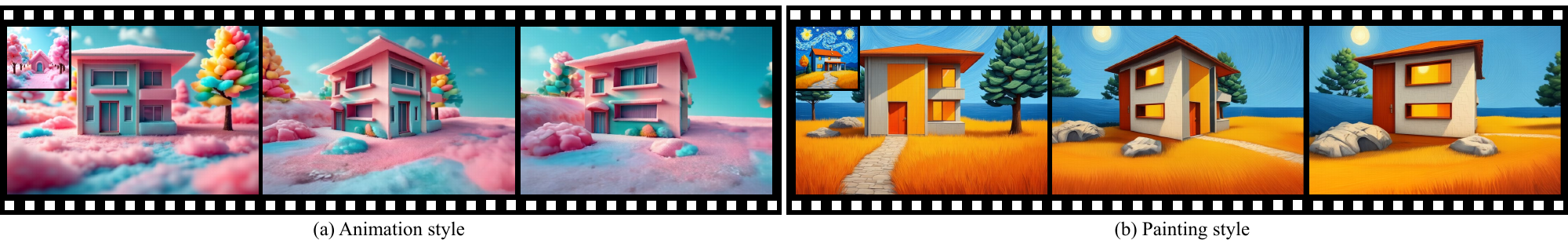}
    \vspace{-6mm}
    \caption{Non-photorealistic generation results. The top-left image is the style reference image.}
    \label{fig:qual_lora_2col}
\end{figure*}

%% file: figures/4dtransition.tex
\begin{figure*}[t]
    \centering
    \includegraphics[width=\linewidth]{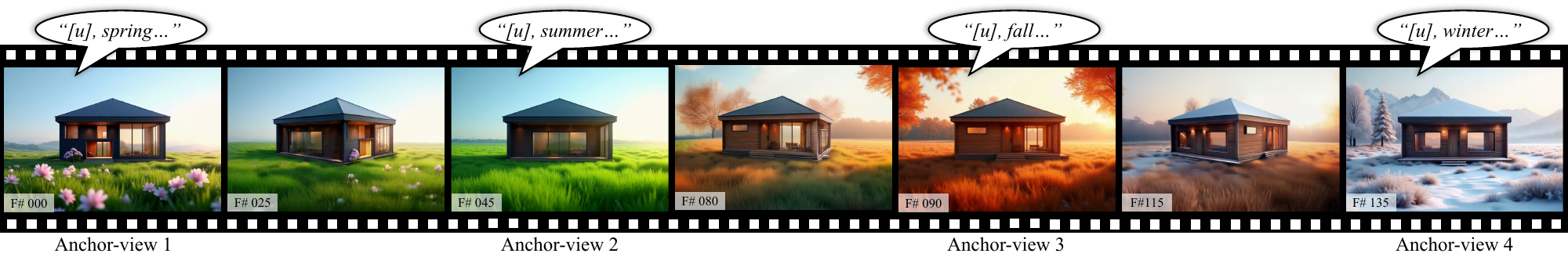}
    \vspace{-6mm}
    \caption{
    Visual effect example showing simultaneous changes in camera motion and temporal context, such as seasonal appearance. Speech bubbles indicate text prompts used in the SAG module, where \textit{[u]} denotes the identifier prompt used in LoRA training.
    }
    \label{fig:4dtransition}
\end{figure*}

%% file: figures/hedofdepth.tex
\begin{figure}
    \centering
    \includegraphics[width=\linewidth]{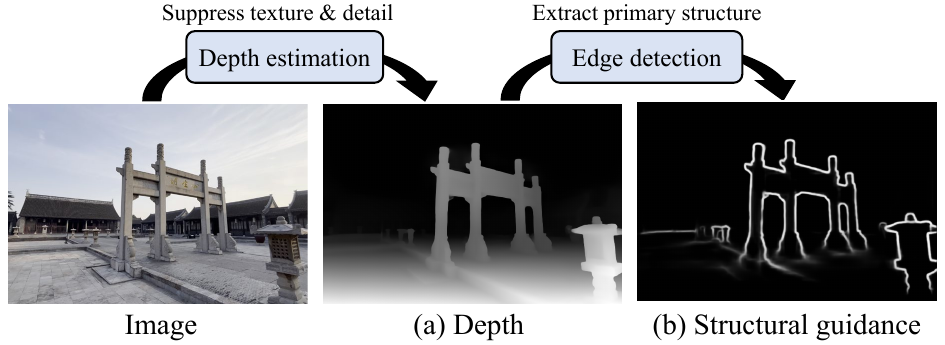}
    \caption{Structural guidance simulation during GGI module training to reduce the domain gap between training and inference. The image is from DL3DV dataset~\cite{dl3dv}.}
    \vspace{-2mm}
    \label{fig:hedofdepth}
\end{figure}

%% file: sections/d_experiments.tex
%
\input{tables/quant_A}

\input{figures/cmp}

\section{Experiments} 
\label{sec:experiments}

\paragraph{Implementation Details}
In the SAG module, the LoRA layers (rank 16) are applied to the first 23 transformer blocks and trained for 400 iterations using a reference image and its HED edge map with the Adam optimizer~\cite{adam}.
In the GGI module, the resolution of the output video is 720$\times$480.
The last frame index $N$ is set to 45, generating 46 frames. 
The module is trained for 1,300 iterations with a batch size of 16 using AdamW optimizer~\cite{adamw}.
During training and inference of the GGI module, we provide a text prompt generated from the first frame using BLIP~\cite{blip}, as the base model, CogVideoX, requires an input text prompt.

\subsection{Video Generation Results}
\cref{fig:qual} shows qualitative results of our method in various scenarios. 
The first example shows that our method works reliably in a simple object-centric scene.
The second one highlights our robustness to dynamic camera motions involving large translations and rotations. 
The third and last columns demonstrate robust performance even in complex spatial transitions, across rooms and hallways, and between indoor and outdoor spaces, respectively.
In addition, \cref{fig:qual_lora_2col} presents results on artistic styles, demonstrating effectiveness in generating non-photorealistic scenes such as animations and paintings.


\cref{fig:4dtransition} shows an example where the style changes over time.
To achieve this, each anchor view is assigned a distinct style via post-prompting with a different seasonal description.
To enable natural scene-style transition, we intentionally omit the replacement operation in the Sparse Appearance-guided Sampling.
Finally, the GGI module smoothly interpolates between distinct style frames, enabling challenging animation.

\input{figures/sag_ablation}

\subsection{Baseline Comparisons}

To validate our method, we compare it with several baselines. 
For the video diffusion-only approach, we compare with VACE~\cite{vace} and a depth-conditioned I2V model, denoted as Depth-I2V.
For VACE inference, we use depth maps as the structural clue, since they yield the best performance compared to other types of clues.
Depth-I2V is trained on DL3DV-10K~\cite{dl3dv} by concatenating depth maps to the latent input, and is initialized from I2V-CogVideoX-5B-1.0.
We also compare with state-of-the-art few-shot reconstruction models, each representing a distinct paradigm: MVSplat360~\cite{mvsplat360} (video diffusion-based), LVSM~\cite{lvsm} (regression-based), and SEVA~\cite{seva} (multi-view diffusion-based).
These models take as input the anchor-view images generated by our SAG module.

\input{tables/ggi_ablation}

To measure visual fidelity, we use PSNR, SSIM, and LPIPS~\cite{lpips}. 
Since ground-truth (GT) intermediate frames are unavailable, we construct pseudo-GT frames by warping the anchor frames $v_0$ and $v_N$ to obtain $v_{0\rightarrow i}$ and $v_{N\rightarrow i}$, where $i$ represents the target frame index. We then composite these warped results, and compute the metrics only on the known regions.
For structural fidelity, we compute PSNR between the GT depth maps rendered from input 3D models, and depth maps estimated by a monocular depth estimator from the corresponding synthesized videos (PSNR-D). To compensate for nonlinear errors and scene-dependent scale variations inherent in monocular depth estimation, we apply histogram equalization before computing PSNR.
For visual quality, we report CLIP-A~\cite{clip-a} and MUSIQ~\cite{musiq} scores.
For style similarity, we measure CLIP image similarity~\cite{clip} (CLIP-I) with the reference style image, as well as Subject Consistency (SC) and Background Consistency (BC) \cite{vbench++}, which compute feature similarity between each frame and both the first and adjacent frames using DINO~\cite{dino} and CLIP, respectively.

For the test dataset, we construct 16 3D models, either manually modeled or sourced from open-source 3D assets~\cite{turbosquid}. The dataset includes 4 object-centric, 2 indoor, 8 outdoor, and 2 indoor-outdoor transition scenes. For each model, we synthesize three different styles using either a distinct reference style image or post-prompting, resulting in a total of 48 generated videos. 

\cref{fig:cmp} shows qualitative comparisons with the baselines. Depth-I2V and VACE generally produce low-quality results with insufficient details, due to the limited generative capability of the video diffusion model.
MVSplat360 often produces severe artifacts due to frequent failures in reconstructing intermediate 3D representations when the distance between anchor views is large.
LVSM generates blurry outputs in regions that require strong generative priors. SEVA often fails under challenging trajectories, mainly due to scale ambiguity arising from its reliance on explicit camera poses (\cref{fig:cmp}c). Even in simpler cases, it deviates significantly from the GT structure (\cref{fig:cmp}d). In contrast, our method achieves higher visual quality and structural fidelity even under challenging conditions. 
\cref{tab:quant} shows quantitative comparisons with the baselines.
Our method achieves the best performance across most metrics and ranks second-best in a few, demonstrating its overall effectiveness.

\input{figures/ggi_ablation}
\input{figures/sag-only}

\subsection{Ablation \& Analysis}
\label{ssec:analysis}

\paragraph{Ablation on SAG Module}
\cref{fig:sag_ablation} presents anchor-view generation results with and without the Sparse Appearance-guided Sampling. The orange boxes indicate corresponding regions in the input geometry. In \cref{fig:sag_ablation}(c), where the guided sampling is not applied, the generated details such as the roof, windows, and facade color patterns significantly deviate from those in the start frame. In contrast, with the guided sampling applied (\cref{fig:sag_ablation}b), these details remain visually consistent, demonstrating the effectiveness of the method.

\paragraph{Ablation on GGI Module}
\cref{fig:ggi_ablation} presents inference results with different GGI modules trained under varying structural conditions.
Without any structural condition, severe distortions frequently occur (\cref{fig:ggi_ablation}a).
Using HED edges directly extracted from the RGB image results in missing details (\cref{fig:ggi_ablation}b).
In contrast, our simulated structural condition, denoted as HED-S, accurately preserves structure and avoids detail loss.
This qualitative observation aligns well with the quantitative comparison in \cref{tab:ggi_ablation}.

\paragraph{Dense View Generation using SAG}
One might wonder whether the SAG module alone could be used to generate intermediate views, rather than relying on the GGI module, since it already produces plausible novel views for the anchor frames.
To investigate this, we compare temporal profiles in \cref{fig:sag-only}, which visualize a fixed 160$\times$20 pixel region over time, comparing our full method and the SAG-only approach.
In the SAG-only setting, frames are generated along the camera trajectory solely using the SAG module.
As shown in the red box, the inherent randomness of the generation process leads to severe flickering and temporal inconsistency.
This highlights the necessity of the GGI module for consistent video synthesis.

\paragraph{Structural Condition for Anchor-view Generation}
\cref{fig:control_type} shows generation results using Flux ControlNet~\shortcite{xlab} with different types of structural conditions, applied to both coarse and detailed geometry.
The Canny-edge condition yields visually monotonous results under coarse geometry due to a mismatch between fine-texture training edges and sparse test-time inputs.
Conversely, the depth condition tends to ignore weak signals in the depth map, making it less effective for guiding detailed geometry. 
In contrast, HED edge conditioning generalizes well to both coarse and detailed cases, as its estimator is trained on sparse, human-annotated edge maps that closely resemble the distribution of 3D-model-derived edges.

\paragraph{Latency}
\cref{tab:latency} shows the latency of each component. After LoRA training, generating a single trajectory takes 197 seconds.

\input{figures/control_type}
\input{tables/latency}

%% file: tables/quant_A.tex
\begin{table*}[t]
    \centering
   \caption{
Quantitative comparisons with Depth-I2V, VACE, and SAG-augmented MVSplat360, LVMS, and SEVA.
    }
    
    \scalebox{0.80}{
\begin{tabular}{l|ccc|c|cc|ccc}
\toprule
     & \multicolumn{3}{c|}{Visual fidelity} & Structural fidelity & \multicolumn{2}{c|}{Visual quality}  & \multicolumn{3}{c}{Style consistency} \\
                     & PSNR$\uparrow$  & SSIM$\uparrow$  & LPIPS$\downarrow$   & PSNR-D$\uparrow$    &  CLIP-A$\uparrow$    & MUSIQ$\uparrow$  & CLIP-I$\uparrow$ & SC$\uparrow$   & BC$\uparrow$   \\ \hline\hline
Depth-I2V            &   -             &    -            &  -              & \textbf{20.696} & 6.136            & 65.240          & 0.787          & 0.864           & 0.920                    \\
VACE~\cite{vace}     &   -             &    -            &  -              & 18.850          & 6.189            & 65.318          & 0.787          & 0.856           & 0.914                    \\
SAG + MVSplat360~\cite{mvsplat360}     & 13.163          & 0.374           & 0.315           & 13.881          & 5.714            & 50.524          & 0.788          & 0.797           & 0.894                    \\
SAG + LVSM~\cite{lvsm}           & \underline{15.103} & \underline{0.472}  & 0.280           & 15.222          & 5.680            & 50.323          & 0.804          & 0.843           & 0.917                    \\
SAG + SEVA~\cite{seva}           & 14.014          & 0.437           & \underline{0.261}  & 16.598          & \textbf{6.782}   & \underline{66.359} & \underline{0.834} & \underline{0.884}  & \underline{0.939}           \\ \rowcolor{yellow!30}
SAG + GGI (Ours)     & \textbf{16.739} & \textbf{0.554}  & \textbf{0.236}  & \underline{19.754} & \underline{6.730}   & \textbf{68.615} & \textbf{0.840} & \textbf{0.891}  & \textbf{0.942}           \\
\bottomrule
\end{tabular}
}
\label{tab:quant}
\end{table*}

%% file: figures/cmp.tex
\begin{figure*}[t]
    \centering
    \includegraphics[width=0.93\linewidth]{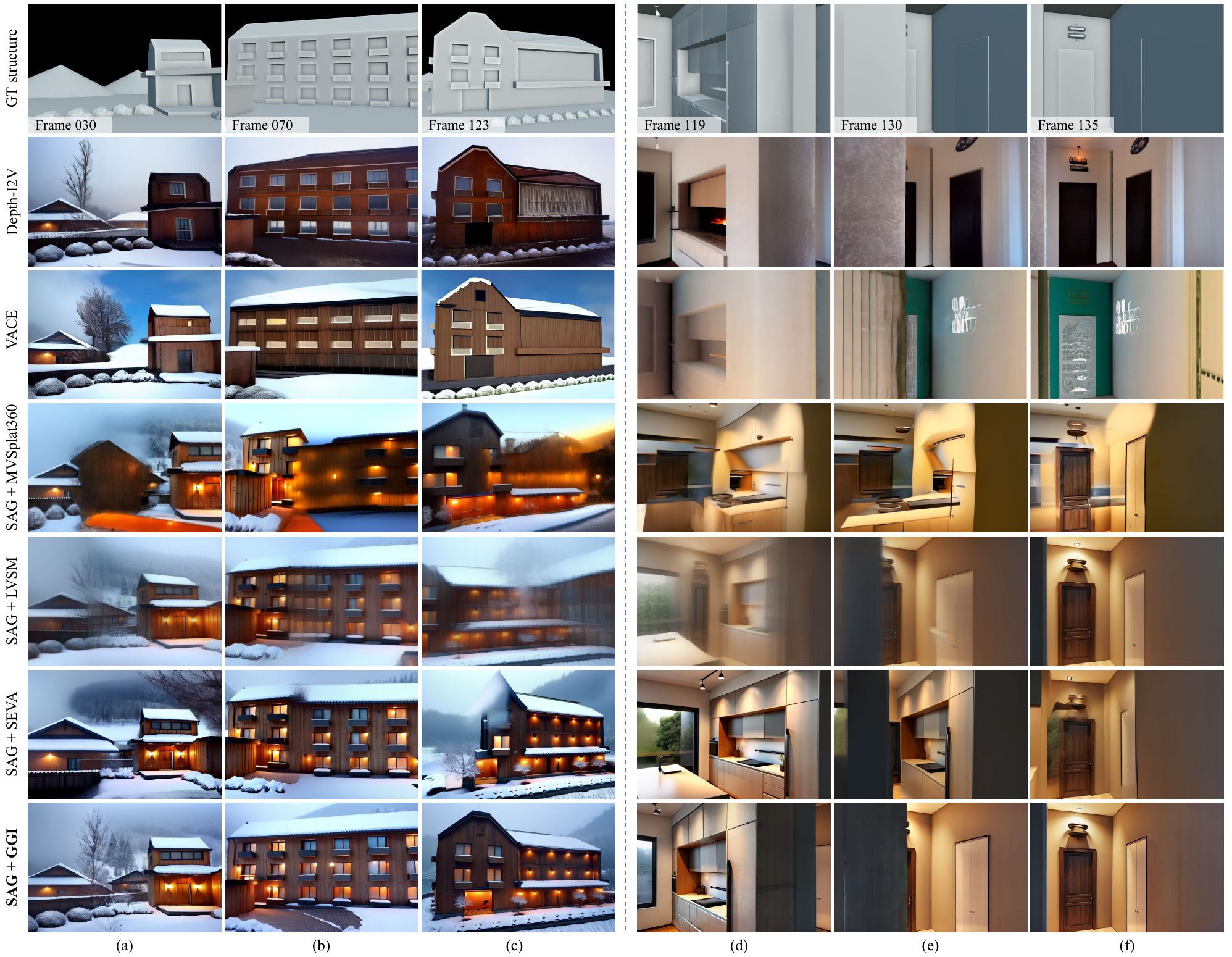}
    \caption{
    Qualitative comparisons with Depth-I2V (a depth-conditioned I2V diffusion model), VACE~\cite{vace}, and SAG-augmented variants of MVSplat360~\cite{mvsplat360}, LVMS~\cite{lvsm}, and SEVA~\cite{seva}. The input geometry of (d-f) is from TurboSquid (\textcopyright 3D LT).
    }   
    \label{fig:cmp}
\end{figure*}

%% file: figures/sag_ablation.tex
\begin{figure}[t]
    \centering
    \includegraphics[width=\linewidth]{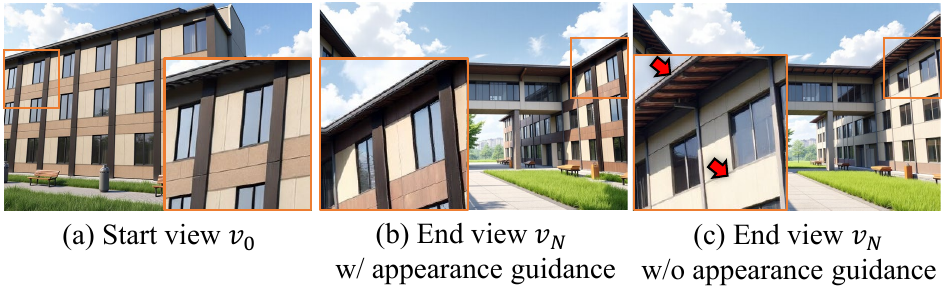}
    \caption{Qualitative results with and without the Sparse Appearance-guided Sampling. Orange boxes indicate the same regions of the input geometry. The input geometry is from TurboSquid (\textcopyright Okhey).
}
    \label{fig:sag_ablation}
\end{figure}

%% file: tables/ggi_ablation.tex
\begin{table}[t]
\centering
\caption{Quantitative comparison of different structural conditions of GGI.}

\scalebox{0.78}{
\begin{tabular}{lcccccc}
\toprule
Condition & PSNR$\uparrow$   & SSIM$\uparrow$   & LPIPS$\downarrow$ & PSNR-D$\uparrow$ & CLIP-A$\uparrow$  & MUSIQ$\uparrow$    \\ \hline\hline
-         & 14.160      	 & 0.417	        & 0.282             & 16.780    	   & 6.622	           & 62.562             \\
HED       & 15.935      	 & 0.545	        & 0.238             & 19.655    	   & 6.580	           & 67.066             \\ \rowcolor{yellow!30}
HED-S     & \textbf{16.739}	 & \textbf{0.554}   & \textbf{0.236}    & \textbf{19.754}  & \textbf{6.730}    & \textbf{68.615}    \\
\bottomrule
\end{tabular}
 }
\label{tab:ggi_ablation}
\end{table}

%% file: figures/ggi_ablation.tex
\begin{figure}[t]
    \centering
    \includegraphics[width=\linewidth]{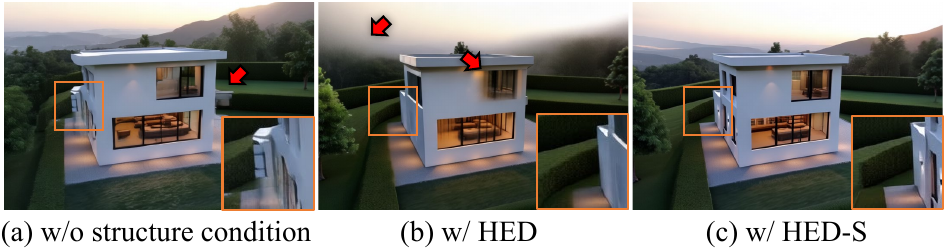}
    \caption{Qualitative comparison of different structural conditions of GGI.}
    \label{fig:ggi_ablation}
\end{figure}

%% file: figures/sag-only.tex
\begin{figure}[t]
    \centering
    \includegraphics[width=0.95\linewidth]{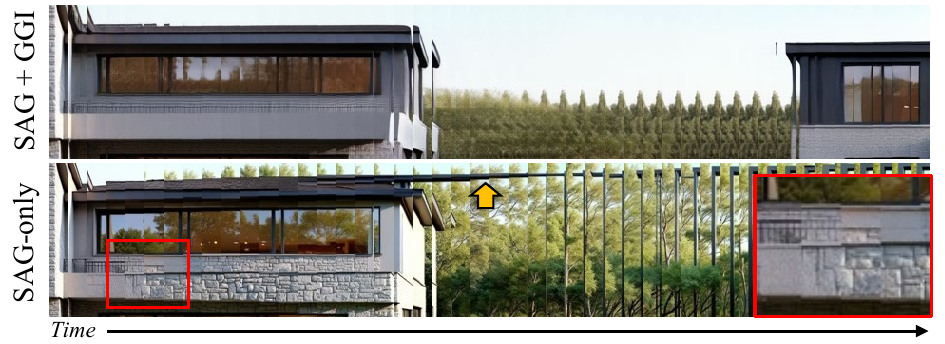}
    \caption{Dense view generation using only the SAG module causes severe flickering (red) and accumulated warping errors (yellow).}
    \label{fig:sag-only}
\end{figure}

%% file: figures/control_type.tex
\begin{figure}[t]
    \centering
    \includegraphics[width=0.95\linewidth]{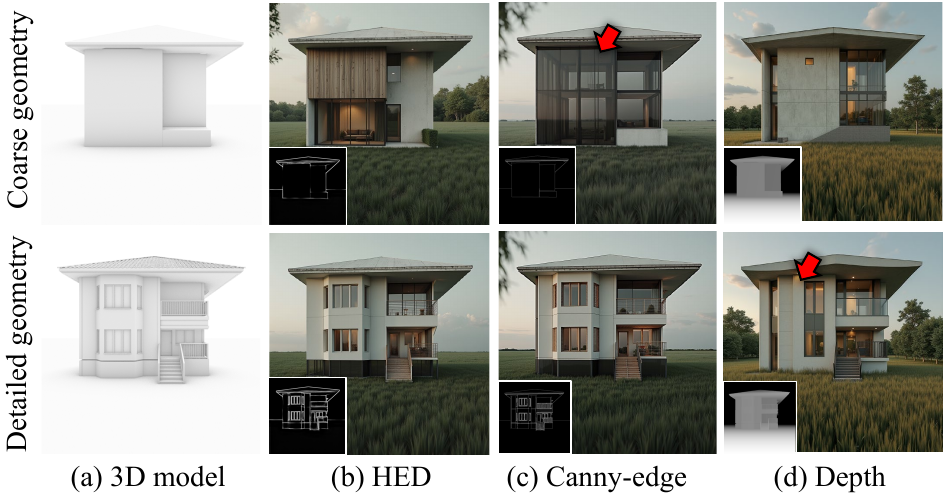}
    \caption{FLUX ControlNet generation results using (b) HED edge map, (c) Canny-edge map and (d) depth map.}
    \label{fig:control_type}
\end{figure}

%% file: tables/latency.tex
\begin{table}[t]
\centering
\caption{
Latency measurements for each process using an A100-80GB GPU.
}
\scalebox{0.78}{
\begin{tabular}{lcccc}
\toprule
        & Preprocessing  & Distribution alignment     & SAG module      & GGI module  \\ \hline\hline
Latency & {12} sec/traj. & {27} min./style & {20} sec./view & {145} sec./traj. \\
\bottomrule
\end{tabular}
}
\label{tab:latency}
\end{table}

%% file: sections/e_conclusion.tex
\section{Conclusion} 
\label{sec:conclusion}

In this paper, we introduce \methodname, a novel framework synthesizing high-quality 3D scene videos given coarse geometry, camera trajectories, and reference images. By combining the complementary strengths of image and video diffusion models through the SAG and GGI modules, our method produces style-consistent, natural, and geometrically faithful videos. Extensive evaluations demonstrate its effectiveness across diverse and challenging scenarios.

\paragraph{Limitations}
\methodname{} does not support real-time interactive camera control.
In addition, temporal inconsistency may occur due to the inherent randomness of diffusion models.
Our method requires LoRA training, which requires a significant amount of computation time, as shown in \cref{tab:latency}.
Addressing these limitations would be an interesting future direction.




